\documentclass[12pt]{iopart}

\usepackage{epsfig}

\begin{document}

\title{A cosmological model in Weyl-Cartan spacetime}
\author{Dirk Puetzfeld \dag and Romualdo Tresguerres \ddag}
\address{\dag Institute for Theoretical Physics, University of Cologne,
  50923 K\"oln, Germany}
\address{\ddag Instituto de matem\'aticas y f\'{\i}sica fundamental, CSIC,
  Serrano 113 bis, 28006 Madrid, Spain}
\ead{dp@thp.uni-koeln.de, ceef310@imaff.cfmac.csic.es}

\begin{abstract}
We present a cosmological model for early stages of the universe on the basis
of a Weyl-Cartan spacetime. In this model, torsion $T^{\alpha}$ and
nonmetricity $Q_{\alpha \beta}$ are proportional to the vacuum
polarization. Extending earlier work of one of us (RT), we discuss the behavior
of the cosmic scale factor and the Weyl 1-form in detail. We show how our model fits into the more general framework of metric-affine gravity (MAG).
\end{abstract}

\submitto{\CQG}
\maketitle

\section{Introduction}

We present a scale invariant model for early stages of the universe.
Spacetime is described by a Weyl-Cartan geometry. The torsion and
nonmetricity become proportional to the Weyl 1-form $Q$. Our starting point
is the work of Tresguerres \cite{Tresguerres}, which will be discussed here
in more detail. The plan of the paper is as follows. In section \ref
{TRESGUERRES_KAPITEL} we derive the field equations and thereby recover the
results found in \cite{Tresguerres}. In sections
\ref{VACUUM_SOLUTION_KAPITEL}--\ref{RADIATIVE_SOLUTION_KAPITEL}, we discuss the behavior of the cosmic scale
factor $S(t)$ and that of the Weyl 1-form $Q$, which governs the
non-Riemannian features of the model. We provide a list of the explicit form
of the surviving curvature pieces for each branch of the solution.
Additionally, we investigate the question of whether the solution exhibits
singularities. In \ref{MAG_KAPITEL} we make contact with the formalism used
in metric-affine gauge theory of gravity as proposed by Hehl \etal in 
\cite{PhysRep}. Finally, in \ref{WEYL_CARTAN_KAPITEL}, we show how the
geometric quantities occurring in MAG have to be constrained in case of a
Weyl-Cartan spacetime.

\section{Cosmology in a Weyl-Cartan spacetime\label{TRESGUERRES_KAPITEL}}

Following the model proposed in \cite{Tresguerres}, we confine ourselves to
a Weyl-Cartan spacetime $Y_{4}$. For a brief introduction into the Weyl-Cartan
subcase of MAG see \ref{WEYL_CARTAN_KAPITEL}. We start start with the following Lagrangian ($R_{\alpha \beta }=W_{\alpha \beta
}+Z_{\alpha \beta }=$ antisymmetric $+$ symmetric part of the curvature): 
\begin{eqnarray}
V &=&\frac{\chi }{2\kappa }R_{\alpha }{}^{\beta }\wedge \eta _{\beta
}\,^{\alpha }+\sum\limits_{I=1}^{6}a_{I}\,\,^{(I)}W_{\alpha }{}^{\beta
}\wedge \,^{\star }R_{\beta }\,^{\alpha }+b\,Z_{\alpha \beta }\wedge
\,^{\star }R^{\beta \alpha }  \label{Tresguerres_lagrangian} \\
&=&\texttt{\rm Einstein-Hilbert}+\texttt{\rm quadratic rotational curvature}  \nonumber \\
&&+\texttt{\rm quadratic strain curvature,}
\end{eqnarray}
with $\chi ,$ $a_{I=1..6},$ $b$ arbitrary constants and $\kappa $ being the
weak coupling constant. Here we make use of the irreducible decomposition of
the curvature as presented in \cite{McCrea}. The matter Lagrangian $L_{{\rm %
mat}}\,$is not explicitly given, but will later on be introduced in a
phenomenological way. From (\ref{Tresguerres_lagrangian}) we calculate the
gauge field excitations (cf. \ref{MAG_KAPITEL}, eqs. (\ref{exications})-(\ref
{matter_currents})):\ 
\begin{eqnarray}
H_{\alpha } &=&0,\quad M^{\alpha \beta }=0,  \label{tresguerrs_excitations1}
\\
H^{\alpha }{}_{\beta } &=&-\frac{\chi }{2\kappa }\eta _{\beta }{}^{\alpha
}-2\sum_{I=1}^{6}a_{I}\,\,^{\star (I)}W_{\beta }{}^{\alpha }-\frac{b}{2}%
\delta _{\beta }^{\alpha }\,^{\star }R^{\gamma }{}_{\gamma }.
\label{tresguerrs_excitations2}
\end{eqnarray}
The canonical energy-momentum is given by 
\begin{eqnarray}
E_{\alpha } &=&e_{\alpha }\rfloor V+\left( e_{\alpha }\rfloor R_{\beta
}{}^{\gamma }\right) \wedge H^{\beta }{}_{\gamma }+\left( e_{\alpha }\rfloor
T^{\beta }\right) \wedge H_{\beta }+\frac{1}{2}\left( e_{\alpha }\rfloor
Q_{\beta \gamma }\right) \wedge M^{\beta \gamma } \\
&=&e_{\alpha }\rfloor V+\left( e_{\alpha }\rfloor R_{\beta }{}^{\gamma
}\right) \wedge H^{\beta }{}_{\gamma }.
\label{canonical_energy_momentum_tresguerres}
\end{eqnarray}
Because of (\ref{tresguerrs_excitations1}), the hypermomentum vanishes 
\begin{equation}
E^{\alpha }{}_{\beta }=-\vartheta ^{\alpha }\wedge H_{\beta }-g_{\beta
\gamma }M^{\alpha \gamma }=0.
\end{equation}
The field equations in a Weyl-Cartan spacetime (cf. \ref{WEYL_CARTAN_KAPITEL}%
) now turn into 
\begin{eqnarray}
-E_{\alpha } &=&\Sigma _{\alpha ,}  \label{first_tresguerres} \\
dH^{\alpha }{}_{\alpha } &=&\Delta ,  \label{second_tresguerres_1} \\
g_{\gamma \lbrack \alpha }DH^{\gamma }\,_{\beta ]} &=&\tau _{\alpha \beta }.
\label{second_tresguerres_2}
\end{eqnarray}
Following \cite{Tresguerres}, we consider a non-massive medium without spin,
i.e. the spin current is assumed\footnote{%
Note that we mark additional assumptions with an ''$A$''.} to vanish $\tau
_{\alpha \beta } \stackrel{A}{=}0$ (this can also be interpreted as a spin current which averages out on
macroscopic scales). Thus, our medium is equipped only with a dilation
current $\Delta $ and eq. (\ref{second_tresguerres_2}) turns into 
\begin{equation}
g_{\gamma \lbrack \alpha }DH^{\gamma }\,_{\beta ]}=0.
\label{second_tresguerre_vanishing_spin_2}
\end{equation}
In addition to (\ref{first_tresguerres}), (\ref{second_tresguerres_1}) and (%
\ref{second_tresguerre_vanishing_spin_2}), one has to consider the Noether
identities (cf. \ref{WEYL_CARTAN_KAPITEL}). They play the role of consistency
conditions, since the matter Lagrangian is not explicitly given. For a vanishing spin current the
second identity (cf. eqs. (\ref{weyl_cartan_second_noether_1}), (\ref
{weyl_cartan_second_noether_2})) turns into 
\begin{eqnarray}
\frac{1}{4}g_{\alpha \beta }\,d\Delta +\vartheta _{(\alpha }\wedge \Sigma
_{\beta )} &=&\sigma _{\alpha \beta },  \label{second_noether_tresguerres_1}
\\
\vartheta _{\lbrack \alpha }\wedge \Sigma _{\beta ]} &=&0.
\label{second_noether_tresguerres_2}
\end{eqnarray}
Taking eq. (\ref{second_tresguerres_1}) into account, (\ref
{second_noether_tresguerres_1}) turns into 
\begin{equation}
\vartheta _{(\alpha }\wedge \Sigma _{\beta )}=\sigma _{\alpha \beta }.
\label{second_noether_treguerres_final_1}
\end{equation}
The first identity (\ref{connection_weyl_cartan_final}) becomes 
\begin{equation}
D\Sigma _{\alpha } \stackrel{\tau_{\alpha \beta}=0\,,\,(\ref{second_noether_treguerres_final_1})}{=}\left( e_{\alpha }\rfloor T^{\beta }\right) \wedge \Sigma _{\beta }-\frac{1}{2}\left( e_{\alpha }\rfloor Q\right) \,\vartheta ^{\beta }\wedge \Sigma
_{\beta }+\frac{1}{4}\left( e_{\alpha }\rfloor R\right) \wedge \Delta .
\label{first_noether_tresguerres_final}
\end{equation}
Thus, we have to solve (\ref{first_tresguerres}), (\ref{second_tresguerres_1}%
), (\ref{second_tresguerre_vanishing_spin_2}), (\ref
{second_noether_tresguerres_2}), (\ref{second_noether_treguerres_final_1}),
and (\ref{first_noether_tresguerres_final}) in order to obtain a solution of
the MAG field equations (\ref{matter})-(\ref{second}).

Next we turn to the description of the matter sources. For a vanishing spin
current in a Weyl-Cartan spacetime (cf. eq. (\ref{Weyl_Cartan_hypermomentum}%
)) the hypermomentum $\Delta _{\alpha \beta }$ becomes proportional to its
trace part, i.e. reduces to the dilation contribution 
\begin{equation}
\Delta _{\alpha \beta }=\frac{1}{4}g_{\alpha \beta }\Delta .
\label{hypermometum_tresguerres_vanishing_spin}
\end{equation}
The trace part (\ref{second_tresguerres_1}) of the second field equation yields 
\begin{equation}
d\Delta =0\quad \Rightarrow \quad \exists \,\,\texttt{\rm 2-form }P:\quad
dP=\Delta .
\end{equation}
Consequently, the dilation current is conserved and there exists a 2-form $P$
being a potential for $\Delta $. We call this form polarization 2-form, the most obvious choice would be 
\begin{equation}
P\stackrel{A}{=}H^{\gamma }{}_{\gamma }=-2b\,^{\star }R=-2b\,^{\star }R^{\gamma }{}_{\gamma
}.  \label{1_ansatz_polarisation_form}
\end{equation}
Of course, this ansatz automatically satisfies (\ref{second_tresguerres_1}).
Guided by (\ref{1_ansatz_polarisation_form}) we assume $P$ to be of the form 
\begin{equation}
P=f(t,r)\,\vartheta ^{\alpha }\wedge \vartheta ^{\beta }.
\label{2_ansatz_polarisation_form}
\end{equation}
So far, $\alpha $ and $\beta $ represent unspecified coframe indices (we fix
them when introducing the coframe), and $f(t,r)$ denotes an arbitrary
function of the time and radial coordinate. After fixing the general form of
the hypermomentum current, we have to specify the energy-momentum 3-form
which appears in (\ref{first_tresguerres}), (\ref
{second_noether_tresguerres_2}), (\ref{second_noether_treguerres_final_1}),
and (\ref{first_noether_tresguerres_final}). Equation (\ref
{second_noether_tresguerres_2}) forces $\Sigma _{\alpha }$ to be symmetric.
Consequently, we choose 
\begin{equation}
\Sigma _{\alpha }\stackrel{A}{=}\Sigma _{\alpha \beta }\,\eta ^{\beta },
\label{energy_momentum_3_form_tresguerres}
\end{equation}
where $\Sigma _{\alpha \beta }=\,$diag$(\mu ,p_{r},p_{t},p_{t})$ and $\mu
=\mu (t),$ $p_{r}=p_{r}(t),$ $p_{t}=p_{t}(t)$ denote the energy density, and
the radial and tangential stresses. Taking (\ref
{second_noether_treguerres_final_1}) into account, one obtains the relation 
\begin{eqnarray}
\sigma ^{\gamma }{}_{\gamma }\stackrel{(\ref{second_noether_treguerres_final_1})}{=}\,\vartheta ^{\gamma }\wedge \Sigma _{\gamma } &=&\vartheta ^{\gamma }\wedge
\left( \Sigma _{\gamma \beta }\eta ^{\beta }\right) =\Sigma _{\gamma \beta
}\,\left( \vartheta ^{\gamma }\wedge \eta ^{\beta }\right)  \nonumber \\
&=&\Sigma _{\gamma \beta }\,\left( \vartheta ^{\gamma }\wedge \,^{\star
}\vartheta ^{\beta }\right) =\Sigma _{\gamma \beta }\left( \left( -1\right)
^{1-1}g^{\gamma \beta }\,\,^{\star }1\right)  \nonumber \\
&=&\Sigma _{\gamma \beta }\,g^{\gamma \beta }\eta =\Sigma ^{\gamma
}{}_{\gamma }\eta =(\mu -p_{r}-2p_{t})\eta .
\label{trace_energy_mometum_tresguerres}
\end{eqnarray}
Now we are going to fix the underlying metrical structure. Following the
standard cosmological model (cf. \cite{KolbTurner}), we take the
Robertson-Walker line element as starting point. 
\begin{equation}
\vartheta ^{\hat{0}}=dt,\quad \vartheta ^{\hat{1}}=\frac{S(t)}{\sqrt{1-kr^{2}%
}}dr,\quad \vartheta ^{\hat{2}}=S(t)\,r\,d\theta ,\quad \vartheta ^{\hat{3}%
}=S(t)\,r\,\sin \theta d\phi ,
\end{equation}
with the line element 
\begin{equation}
ds^{2}\stackrel{A}{=}\vartheta ^{\hat{0}}\otimes \vartheta ^{\hat{0}}-\vartheta ^{\hat{1}}\otimes
\vartheta ^{\hat{1}}-\vartheta ^{\hat{2}}\otimes \vartheta ^{\hat{2}%
}-\vartheta ^{\hat{3}}\otimes \vartheta ^{\hat{3}}.
\label{tresguerres_line_element}
\end{equation}
As usual, $S(t)$ denotes the cosmic scale factor and $k=-1,0,1$ determines
whether the spatial sections are of constant negative, vanishing or positive
curvature. Thus, we look for spherically symmetric solutions. This choice
fixes the indices in (\ref{2_ansatz_polarisation_form}) to be $\alpha =\hat{2%
}$ and $\beta =\hat{3}$. Additionally, we impose another constraint on the
so called polarization function $f(t,r)$ in (\ref{2_ansatz_polarisation_form}%
) by allowing only for functions which depend on the time coordinate, i.e.$%
\,f(r,t)\stackrel{A}{\rightarrow}\frac{\xi (t)}{S^{2}(t)}$. Consequently, our ansatz made in (\ref
{2_ansatz_polarisation_form}) turns into 
\begin{equation}
P=\frac{\xi (t)}{S^{2}(t)}\vartheta ^{\hat{2}}\wedge \vartheta ^{\hat{3}}%
\texttt{\rm ,}  \label{3_ansatz_polarisation_form}
\end{equation}
where $\xi (t)$ represents the new arbitrary polarization function. Equation
(\ref{3_ansatz_polarisation_form}) yields the form of the dilation current $%
\Delta $ to be\footnote{%
Here we made use of $\,^{\cdot }$ $:=\frac{\partial }{\partial t}$ and $\eta
^{\alpha }:=\,^{\star }\vartheta ^{\alpha }.$} 
\begin{eqnarray}
\Delta &=&dP=\frac{2\xi \sqrt{1-kr^{2}}}{S^{3}r}\vartheta ^{\hat{1}}\wedge
\vartheta ^{\hat{2}}\wedge \vartheta ^{\hat{3}}+\frac{\dot{\xi}}{S^{2}}%
\vartheta ^{\hat{0}}\wedge \vartheta ^{\hat{2}}\wedge \vartheta ^{\hat{3}}
\label{dilation_current_tresguerres_explicit} \\
&=&\frac{2\xi \sqrt{1-kr^{2}}}{S^{3}r}\eta ^{\hat{0}}+\frac{\dot{\xi}}{S^{2}}%
\eta ^{\hat{1}}.
\end{eqnarray}
The next thing to come up is a proper ansatz for the torsion 2-form $T^{\alpha }$%
. We choose the torsion to be proportional to its vector piece$\,T^{\alpha
}\sim \,^{(2)}T^{\alpha }$ which is proportional to the Weyl 1-form 
\begin{equation}
T^{\alpha }\stackrel{A}{=}\frac{1}{2}Q\wedge \vartheta ^{\alpha }.  \label{ansatz_torsion_tresguerres}
\end{equation}
We are now going to calculate the Weyl 1-form from the trace part of the
second field equation (\ref{second_tresguerres_1}). Using (\ref
{dilation_current_tresguerres_explicit}), eq. (\ref{second_tresguerres_1})
turns into 
\begin{equation}
dH^{\gamma }{}_{\gamma }=\Delta =dP.
\end{equation}
From (\ref{tresguerrs_excitations2}) we subsequently deduce 
\begin{equation}
dH^{\gamma }{}_{\gamma }=-2b\,d^{\star }R\stackrel{{\rm (Weyl-Cartan)}}{=}-4b\,d^{\star }dQ.
\end{equation}
Putting the last two equations together we obtain 
\begin{eqnarray}
-4bd\,^{\star }dQ &=&dP,  \label{weyl_one_form_solution_treguerres} \\
\Rightarrow \,^{\star }dQ &=&-\frac{P}{4b}\Rightarrow dQ=\frac{\xi }{4bS^{2}}%
\vartheta ^{\hat{0}}\wedge \vartheta ^{\hat{1}}, \\
\Rightarrow \quad \,Q &=&-\frac{\xi }{4bS}\left(\int dr\frac{1}{\sqrt{1-kr^{2}}}\right)\vartheta ^{\hat{0}}+\texttt{\rm exact contribution.}
\end{eqnarray}
Neglecting the trivial exact contribution, we obtain the following explicit
expressions for $Q,$ which depend on the sign of the spatial curvature: 
\begin{eqnarray}
Q &=-\frac{\xi (t)}{4bS(t)}\,\texttt{\rm arcsinh}\left( r\right) \,\vartheta ^{%
\hat{0}}\quad &\texttt{\rm for }k=-1, \\
Q &=-\frac{\xi (t)}{4bS(t)}\,r\,\vartheta ^{\hat{0}}\quad &\texttt{\rm for }k=0, \\
Q &=-\frac{\xi (t)}{4bS(t)}\,\texttt{\rm arcsin}\left( r\right) \,\vartheta ^{%
\hat{0}}\quad &\texttt{\rm for }k=1.
\end{eqnarray}
We calculate the field equations resulting from the second Noether identity
in (\ref{second_noether_treguerres_final_1}). We make use of computer
algebra and obtain two independent equations, namely 
\begin{eqnarray}
\frac{d}{dt}\left( \mu S^{4}+\frac{\xi ^{2}}{16b}\right) &=&\frac{1}{4}%
\left( \mu -p_{r}-2p_{t}\right) \,\frac{dS^{4}}{dt},
\label{sec_noether_explicit1_final} \\
p_{r}-p_{t} &=&\frac{\xi ^{2}}{8bS^{4}}.  \label{sec_noether_explicit2_final}
\end{eqnarray}
We make use of eq. (\ref{sec_noether_explicit2_final}) and rewrite the
energy-momentum trace 
\begin{equation}
\Sigma ^{\gamma }{}_{\gamma }=\mu -p_{r}-2p_{t}=\mu -3p_{t}-\frac{\xi ^{2}}{%
8bS^{4}}.  \label{energy_momentum_trace_tresguerres}
\end{equation}
Let us now inspect the remaining field equations, i.e. the antisymmetric
part of the second equation (\ref{second_tresguerre_vanishing_spin_2}),
called the spin equation, and the first field equation (\ref
{first_tresguerres}). By use of computer algebra, we investigate eq. (\ref
{second_tresguerre_vanishing_spin_2}) and obtain one independent equation,
namely 
\begin{equation}
\left( a_{4}+a_{6}\right) \bigg(\stackrel{...}{S}S^{2}+\ddot{S}\dot{S}S-2 \dot{S}^{3}-2k \dot{S}\bigg)=0.  \label{second_antisymmetric_explicit_tresguerres}
\end{equation}
Thus, for $a_{6}\neq -a_{4}$ and after some algebra, this equation turns
into 
\begin{equation}
 \frac{d}{dt}\left( \frac{\ddot{S}}{S}+\left( \frac{\dot{S}}{S%
}\right) ^{2}+\frac{k}{S^{2}}\right) = 0.  \label{tresguerre_eq_4_13}
\end{equation}
Hereby we recovered one of the field equations given in \cite
{Tresguerres} eq. (3.7). Now we draw our attention to the first field
equation. Eq. (\ref{first_tresguerres}) yields three independent equations,
namely 
\begin{eqnarray}
\chi \left( \left( \frac{\dot{S}}{S}\right) ^{2}+\frac{k}{S^{2}}\right)
-\kappa \left( a_{4}+a_{6}\right) \left( \left( \frac{\ddot{S}}{S}\right)
^{2}-\left( \left( \frac{\dot{S}}{S}\right) ^{2}+\frac{k}{S^{2}}\right)
^{2}\right)   \nonumber \\
=\frac{\kappa }{3}\left( \mu +\frac{\xi ^{2}}{16bS^{4}}\right) , 
\label{first_field_eq_explicit_1_tresguerres}
\end{eqnarray}
\begin{eqnarray}
\chi \left( 2\frac{\ddot{S}}{S}+\left( \frac{\dot{S}}{S}\right) ^{2}+\frac{k%
}{S^{2}}\right) +\kappa \left( a_{4}+a_{6}\right) \left( \left( \frac{\ddot{S%
}}{S}\right) ^{2}-\left( \left( \frac{\dot{S}}{S}\right) ^{2}+\frac{k}{S^{2}}%
\right) ^{2}\right) &&  \nonumber \\
=-\kappa \left( p_{r}-\frac{\xi ^{2}}{16bS^{4}}\right) , &&
\label{first_field_eq_explicit_2_tresguerres}
\end{eqnarray}
\begin{eqnarray}
\chi \left( 2\frac{\ddot{S}}{S}+\left( \frac{\dot{S}}{S}\right) ^{2}+\frac{k%
}{S^{2}}\right) +\kappa \left( a_{4}+a_{6}\right) \left( \left( \frac{\ddot{S%
}}{S}\right) ^{2}-\left( \left( \frac{\dot{S}}{S}\right) ^{2}+\frac{k}{S^{2}}%
\right) ^{2}\right) &&  \nonumber \\
=-\kappa \left( p_{t}+\frac{\xi ^{2}}{16bS^{4}}\right) . &&
\label{first_field_eq_explicit_3_tresguerres}
\end{eqnarray}
Comparison with the results in \cite{Tresguerres} reveals the equivalence of (\ref
{first_field_eq_explicit_1_tresguerres}), (\ref
{first_field_eq_explicit_3_tresguerres}) and eqs. (3.8),(3.9) of \cite
{Tresguerres}. We make use of some algebra in order to write (\ref
{first_field_eq_explicit_1_tresguerres})-(\ref
{first_field_eq_explicit_3_tresguerres}) in a more convenient form. By
adding (\ref{first_field_eq_explicit_1_tresguerres}) and (\ref
{first_field_eq_explicit_3_tresguerres}) we obtain
\begin{eqnarray}
\chi \left( 2\frac{\ddot{S}}{S}+2\left( \frac{\dot{S}}{S}\right) ^{2}+2\frac{%
k}{S^{2}}\right) &=&\frac{\kappa }{3}\left( \mu +\frac{\xi ^{2}}{16bS^{4}}%
\right) -\kappa \left( p_{t}+\frac{\xi ^{2}}{16bS^{4}}\right)  \nonumber \\
\Leftrightarrow 2\chi \left( \frac{\ddot{S}}{S}+\left( \frac{\dot{S}}{S}%
\right) ^{2}+\frac{k}{S^{2}}\right) &=&\frac{\kappa }{3}\left( \mu -3p_{t}-%
\frac{\xi ^{2}}{8bS^{4}}\right) .
\label{addierte_Feldgleichungen_1_tresguerres}
\end{eqnarray}
Subtracting (\ref{first_field_eq_explicit_3_tresguerres}) from (\ref
{first_field_eq_explicit_1_tresguerres}) yields 
\begin{eqnarray}
2\chi \frac{\ddot{S}}{S}+2\kappa \left( a_{4}+a_{6}\right) \left( \left( 
\frac{\ddot{S}}{S}\right) ^{2}-\left( \left( \frac{\dot{S}}{S}\right) ^{2}+%
\frac{k}{S^{2}}\right) ^{2}\right)\nonumber \\
=-\frac{\kappa }{3}\left( \mu +3p_{t}+\frac{\xi ^{2}}{4bS^{4}}\right) .
\label{subtrahierte_Feldgleichungen_2_tresguerres}
\end{eqnarray}
Closer examination of (\ref{addierte_Feldgleichungen_1_tresguerres}) and (%
\ref{tresguerre_eq_4_13}) leads to
\begin{eqnarray}
0&\stackrel{(\ref{tresguerre_eq_4_13})}{=}&\frac{d}{dt}\left(
  \frac{\ddot{S}}{S}+\left( \frac{\dot{S}}{S}\right)
  ^{2}+\frac{k}{S^{2}}\right)
\stackrel{(\ref{addierte_Feldgleichungen_1_tresguerres})}{=}\frac{\kappa
  }{3}\frac{d}{dt}\left( \mu -3p_{t}-\frac{\xi ^{2}}{8bS^{4}}\right) \\
&\stackrel{(\ref{energy_momentum_trace_tresguerres})}{=}&\frac{\kappa }{3}\frac{d\Sigma ^{\gamma }{}_{\gamma }}{dt}\quad
\Rightarrow \quad \Sigma ^{\gamma }{}_{\gamma }=\mu -p_{r}-2p_{t}=\texttt{\rm constant}=:\Xi .  \label{trace_ist_costant}
\end{eqnarray}
Thus, the trace of the energy-momentum turns out to be a constant. We will
now investigate the consequences of a vanishing energy-momentum trace $\Xi $%
. In order to extract some information from the assumption $\Xi \stackrel{A}{=}0,$ we rewrite eq. (\ref{sec_noether_explicit1_final}) by making use of the
following definitions: 
\begin{equation}
\tilde{\mu}:=\mu +\frac{\xi ^{2}}{16bS^{4}},\quad \quad \tilde{p}:=\frac{1}{3%
}\tilde{\mu}.  \label{effective_energy_and_pressure_definition}
\end{equation}
We call $\tilde{\mu}$ and $\tilde{p}$ effective energy and pressure,
respectively. After some algebra, we find that (\ref
{sec_noether_explicit1_final}) is equivalent to 
\begin{equation}
d\left( \tilde{\mu}S^{3}\right) =-\left( \tilde{p}-\frac{\Xi }{3}\right)
dS^{3}.  \label{tresguerres_eq_5_4}
\end{equation}
This equation has the character of a thermodynamical relation and shows that
in case of $\Xi =0$ the constant contribution to the effective pressure on
the r.h.s.\thinspace vanishes. Consequently, the following relations hold in
case of $\Xi =0$: 
\begin{eqnarray}
\mu &\stackrel{(\ref{trace_ist_costant})}{=}&p_{r}+2p_{t}  \label{energy_for_xi_zero} \\
&\stackrel{(\ref{sec_noether_explicit2_final})}{\Rightarrow}&p_{r}=\frac{1}{3}\left(
  \mu +\frac{\xi ^{2}}{4bS^{4}}\right) \quad \Rightarrow
\quad p_{t}=\frac{1}{3}\left( \mu -\frac{\xi ^{2}}{8bS^{4}}\right) .
\label{stresses_for_xi_zero_1}
\end{eqnarray}
From the second Noether identity in (\ref{sec_noether_explicit1_final}) we
gain $\mu =\mu (t)$ as a function of the polarization function $\xi (t)$ and
the scale factor as follows: 
\begin{equation}
\frac{d}{dt}\left( \mu S^{4}+\frac{\xi ^{2}}{16b}\right) =0\quad \Rightarrow
\quad \mu =-\frac{\xi ^{2}}{16bS^{4}}+\frac{\mu _{0}}{S^{4}},
\label{mue_als_funktion_von_xi_und_s}
\end{equation}
where $\mu _{0}$ represents an integration constant. The stresses in (\ref
{stresses_for_xi_zero_1}) take the form 
\begin{equation}
p_{r}=\frac{\mu _{0}}{3S^{4}}+\frac{\xi ^{2}}{16bS^{4}},\quad \quad p_{t}=%
\frac{\mu _{0}}{3S^{4}}-\frac{\xi ^{2}}{16bS^{4}},
\label{stresses_for_xi_zero_2}
\end{equation}
i.e. are equal if the polarization function $\xi (t)$ vanishes. Once again,
we consider the field equations. From eq. (\ref{tresguerre_eq_4_13}) follows 
\begin{equation}
\frac{\ddot{S}}{S}+\left( \frac{\dot{S}}{S}\right) ^{2}+\frac{k}{S^{2}}=%
\texttt{\rm const}:=\frac{2}{3}\Lambda .  \label{induce_cosmo_constant}
\end{equation}
Note that we are free to choose the emerging constant. Comparison with the
classical Friedman equations reveals that this constant plays a role similar
to that of the cosmological constant. Thus, we call $\Lambda $ \textit{%
induced} cosmological constant. For vanishing trace the field equation (\ref
{addierte_Feldgleichungen_1_tresguerres}) becomes 
\begin{equation}
2\chi \left( \frac{\ddot{S}}{S}+\left( \frac{\dot{S}}{S}\right) ^{2}+\frac{k%
}{S^{2}}\right) =0,  \label{addierte_feldgleichung_fuer_spur_null}
\end{equation}
by insertion of (\ref{induce_cosmo_constant}) we arrive at 
\begin{equation}
\chi \Lambda =0.  \label{addierte_feldgl_final_fuer_spur_null}
\end{equation}
We make use of (\ref{mue_als_funktion_von_xi_und_s}), (\ref
{stresses_for_xi_zero_2}), and (\ref{induce_cosmo_constant}) and rewrite the
remaining field equation (\ref{subtrahierte_Feldgleichungen_2_tresguerres})
as 
\begin{equation}
\chi \frac{\ddot{S}}{S}+\frac{4\kappa \Lambda }{3}\left( a_{4}+a_{6}\right)
\left( \frac{\ddot{S}}{S}-\frac{\Lambda }{3}\right) =-\frac{\kappa \mu _{0}}{%
3S^{4}},  \label{subtrahierte_feldgleichung_final_fuer_xi_null}
\end{equation}
in accordance with the result obtained in \cite{Tresguerres}
eq. (3.14).

Before we discuss possible solutions of the model under consideration, we
will shortly collect all assumptions made up to here in table \ref{tabelle_1}.
\begin{table}
\caption{Assumptions made up to this point.}
\label{tabelle_1}
\begin{indented}
\item[]\begin{tabular}{@{}lll}
\br
Ansatz/assumption & Resulting quantity/equations & Equation number \\ 
\mr
$\tau _{\alpha \beta }=0$ & $g_{\gamma \lbrack \alpha }DH^{\gamma }{}_{\beta
]}=0$ & (\ref{second_tresguerre_vanishing_spin_2}) \\ 
$P=\frac{\xi }{S^{2}}\vartheta ^{\hat{2}}\wedge \vartheta ^{\hat{3}}$ & $%
\Delta ,Q,$ $dH^{\gamma }{}_{\gamma }\equiv 0$ & (\ref
{hypermometum_tresguerres_vanishing_spin}),(\ref
{weyl_one_form_solution_treguerres}),(\ref{second_tresguerres_1}) \\ 
$T^{\alpha }=\frac{1}{2}Q\wedge \vartheta ^{\alpha }$ & Affects form of the
connection & (\ref{Weyl_Cartan_Konnektion}) \\ 
$\left( a_{4}+a_{6}\right) \neq 0$ & Affects form of the spin equation & (%
\ref{tresguerre_eq_4_13}) \\
$\Xi =0=\Sigma ^{\gamma }{}_{\gamma }$ & $\mu $, $\,p_{r}$, $\,p_{t}$, form
of the field eqs, $\Lambda $ & (\ref{mue_als_funktion_von_xi_und_s}),(\ref
{stresses_for_xi_zero_2}),(\ref{induce_cosmo_constant}) \\ 
$\Lambda $ & Affects form of the field eqs. & (\ref
{addierte_feldgleichung_fuer_spur_null})-(\ref
{subtrahierte_feldgleichung_final_fuer_xi_null}) \\ 
\br
\end{tabular}
\end{indented}
\end{table}
The final form of the field equations is given by (\ref
{induce_cosmo_constant}), (\ref{addierte_feldgl_final_fuer_spur_null}), (\ref
{subtrahierte_feldgleichung_final_fuer_xi_null}), i.e. 
\begin{eqnarray}
\frac{\ddot{S}}{S}+\left( \frac{\dot{S}}{S}\right) ^{2}+\frac{k}{S^{2}} &=&%
\frac{2}{3}\Lambda ,  \label{1_final} \\
\chi \Lambda &=&0,  \label{2_final} \\
\chi \frac{\ddot{S}}{S}+\frac{4\kappa \Lambda }{3}\left( a_{4}+a_{6}\right)
\left( \frac{\ddot{S}}{S}-\frac{\Lambda }{3}\right) &=&-\frac{\kappa \mu _{0}%
}{3S^{4}}.  \label{3_final}
\end{eqnarray}

\section{Vacuum solution\label{VACUUM_SOLUTION_KAPITEL}}

\begin{figure}
  \begin{center}
    \epsfig{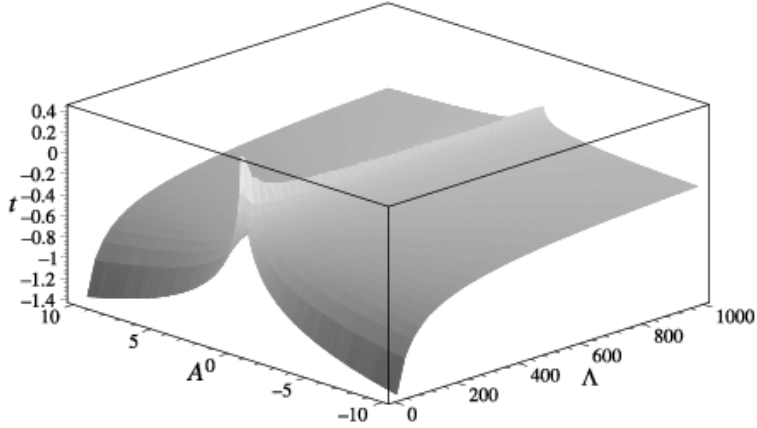}
     \caption[Singularities of the vacuum solution for $k=-1$]{Plot of the times
       $t=t_{\rm div}$ at which the vacuum solution becomes singular in case of $k=-1$.}
     \label{figInvariantDivergence}
   \end{center}
\end{figure}

We are now going to solve the system (\ref{1_final})-(\ref{3_final}). We
start by putting $\chi =0$, i.e. the Einstein-Hilbert term in the Lagrangian
(\ref{Tresguerres_lagrangian}) is assumed to vanish. Since the
Einstein-Hilbert term is supposed to describe physics at low energies, we
expect the upcoming solution to be valid only in very early stages of the
universe, i.e. at high energies. We note that our ansatz fulfills (\ref
{2_final}) and try to solve (\ref{1_final}) for nonvanishing $\Lambda $. 
We can verify that 
\begin{equation}
S(t)=A\,e^{\sqrt{\frac{\Lambda }{3}}t}+\frac{3k}{4A\Lambda }\,e^{-\sqrt{%
\frac{\Lambda }{3}}t}\quad \quad \texttt{\rm , with }A=\texttt{\rm const,}
\label{vacuum_scale_factor}
\end{equation}
satisfies eq. (\ref{1_final})\footnote{$[A]=$ length.}. Insertion of (\ref
{vacuum_scale_factor}) into (\ref{3_final}) with $\chi =0$, yields 
\begin{equation}
\frac{4\kappa \Lambda }{3}\left( a_{4}+a_{6}\right) \left( \frac{\ddot{S}}{S}%
-\frac{\Lambda }{3}\right) =-\frac{\kappa \mu _{0}}{3S^{4}}\quad \rightarrow
\quad \kappa \mu _{0}\Lambda ^{4}A^{4}e^{4\sqrt{\frac{\Lambda }{3}}t}=0.
\label{s_t_in_final_3}
\end{equation}
Thus, we have to choose $\mu _{0}=0$ in order to fulfill (\ref{3_final}).
From eq. (\ref{mue_als_funktion_von_xi_und_s}) we derive the final
expressions of the energy and stresses 
\begin{equation}
^{{\rm vac}}\mu =-\frac{\xi ^{2}}{16bS^{4}}\quad \Rightarrow \quad \,^{{\rm vac}}\tilde{\mu}=0\quad 
\stackrel{(\ref{stresses_for_xi_zero_2})}{\Rightarrow}\quad \,^{{\rm vac}}p_{r}=-\,^{{\rm vac}}p_{t}=-\,^{{\rm vac}}\mu .
\label{stresses_and_energy_for_vacuum_case_final}
\end{equation}
In view of (\ref{stresses_and_energy_for_vacuum_case_final}), it becomes
clear why this branch of our model is the called vacuum solution. Note that in
contrast to the standard result, the vacuum energy $^{{\rm vac}}\mu $ is
not a constant and proportional to the polarization function $\xi (t)$.
Finally, the Weyl 1-form reads 
\begin{eqnarray}
Q &=-\frac{\xi (t)}{4bA\,e^{\sqrt{\frac{\Lambda }{3}}t}-\frac{3b}{\Lambda A}%
\,e^{-\sqrt{\frac{\Lambda }{3}}t}}\,\,\texttt{\rm arcsinh}(r)\,\vartheta ^{\hat{0}%
}\quad \quad &\texttt{\rm for }k=-1\texttt{\rm ,}  \label{weyl_vacuum_k_minus} \\
Q &=-\frac{\xi (t)r}{4bA\,e^{\sqrt{\frac{\Lambda }{3}}t}}\,\vartheta ^{\hat{%
0}}\quad \quad &\texttt{\rm for }k=0\texttt{\rm ,}  \label{weyl_vacuum_k_null} \\
Q &=-\frac{\xi (t)}{4bA\,e^{\sqrt{\frac{\Lambda }{3}}t}+\frac{3b}{\Lambda A}%
\,e^{-\sqrt{\frac{\Lambda }{3}}t}}\,\,\texttt{\rm arcsin}(r)\vartheta ^{\hat{0}%
}\quad \quad &\texttt{\rm for }k=1\texttt{\rm .}  \label{weyl_vacuum_k_plus}
\end{eqnarray}
\newline
We observe that the only surviving pieces of the curvature are $%
^{(6)}W_{\alpha \beta }\,$\ and $^{(4)}Z_{\alpha \beta }$. Their
nonvanishing components read as follows: 
\begin{eqnarray}
^{(6)}W^{\alpha \beta } &=&\frac{\Lambda }{3}\vartheta ^{\alpha }\wedge
\vartheta ^{\beta },  \label{vacuum_rotational_curvature} \\
^{(4)}Z_{\hat{0}\hat{0}} &=&\frac{2A^{2}\Lambda ^{2}e^{2\sqrt{\frac{\Lambda 
}{3}}t}\xi (t)}{b\left( 16A^{4}\Lambda ^{2}e^{4\sqrt{\frac{\Lambda }{3}}%
t}+24kA^{2}\Lambda e^{2\sqrt{\frac{\Lambda }{3}}t}+9\right) }\vartheta ^{%
\hat{0}}\wedge \vartheta ^{\hat{1}}  \nonumber \\
&=&\,-^{(4)}Z_{\hat{1}\hat{1}}=\,-^{(4)}Z_{\hat{2}\hat{2}}=-\,^{(4)}Z_{\hat{3%
}\hat{3}}\texttt{\rm .}  \label{vacuum_strain_curvature}
\end{eqnarray}
\newline
Note that the results collected in (\ref{vacuum_rotational_curvature}) and (%
\ref{vacuum_strain_curvature}) are valid for $k=-1,0,1$. The torsion and
nonmetricity read 
\begin{eqnarray}
^{(2)}T^{\hat{0}} &=0,\quad \quad \quad ^{(2)}T^{a}&=\frac{1}{2}Q\wedge \vartheta ^{a},
\\
^{(4)}Q_{\hat{0}\hat{0}} &=Q=-\,^{(4)}Q_{\hat{1}\hat{1}}&=\,-\,^{(4)}Q_{\hat{%
2}\hat{2}}=\,-\,^{(4)}Q_{\hat{3}\hat{3}}.\,\,
\end{eqnarray}
We continue with the calculation of the curvature invariant $\,^{\star
}\left( R_{\alpha \beta }\wedge \,^{\star }R^{\alpha \beta }\right) $ in
order to answer the question of whether our solution possesses essential
singularities or not. In table \ref{tabelle_2}, we display parameter values and
times at which an essential singularity emerges. In the case of $t_{{\rm div}}$
these are the epochs of the universe at which the invariant diverges. Note
that not listed variables and parameters are allowed to take arbitrary
values.
\begin{table}
\caption{Parameter values and times with essential singularities (vacuum solution).}
\label{tabelle_2}
\begin{indented}
\item[]
\begin{tabular}{@{}ll}
\br
$k=-1$ & $\{t_{{\rm div}}=\frac{\sqrt{3}}{2}\frac{\ln \left( \frac{3}{%
A^{2}\Lambda }\right) }{\sqrt{\Lambda }}\}$ or $\{b=0\}$ \\
$k=0$ & $\{A=0\}$ or$\,\,\,\{b=0\}$ \\
$k=1$ & $\{t_{{\rm div}}=\frac{\sqrt{3}}{2}\frac{\ln \left( -\frac{3}{%
A^{2}\Lambda }\right) }{\sqrt{\Lambda }}\}\,$or $\{b=0\}$ \\
\br
\end{tabular}
\end{indented}
\end{table}
In figure \ref{figInvariantDivergence} we display the function
$t=t_{{\rm div}}(A,\Lambda )$ in the case of $k=-1$. 
As can be read of from there, our solution exhibits no singularity as long as $A$ vanishes.
In the light of (\ref{vacuum_scale_factor}) this choice is only allowed if $%
k=0$. Consequently, our solution always exhibits a singularity after a
finite time as long as $\Lambda \in \,]0,\infty \lbrack $ (negative values
of $\Lambda $ can be ruled out since they lead to a complex valued scale
factor $S(t)$). Reinsertion of the expression for $t_{{\rm div}}(A,\Lambda )$ into the solution for the cosmic scale factor, as stated in (\ref
{vacuum_scale_factor}), reveals that we are dealing with a point singularity
at the origin of the universe\footnote{Note that this statement holds for $k=-1,1.$}, i.e. $S(t_{{\rm div}})=0$.
Thus, this type of singularity is similar to those known from the Riemannian
case. In case of a vanishing $k$ there is no singularity since $b=0$ leads
to a constraint in our Lagrangian and $A=0$ corresponds to the unphysical
solution $S(t)\equiv 0$. 
\begin{figure}
  \begin{center}
    \epsfig{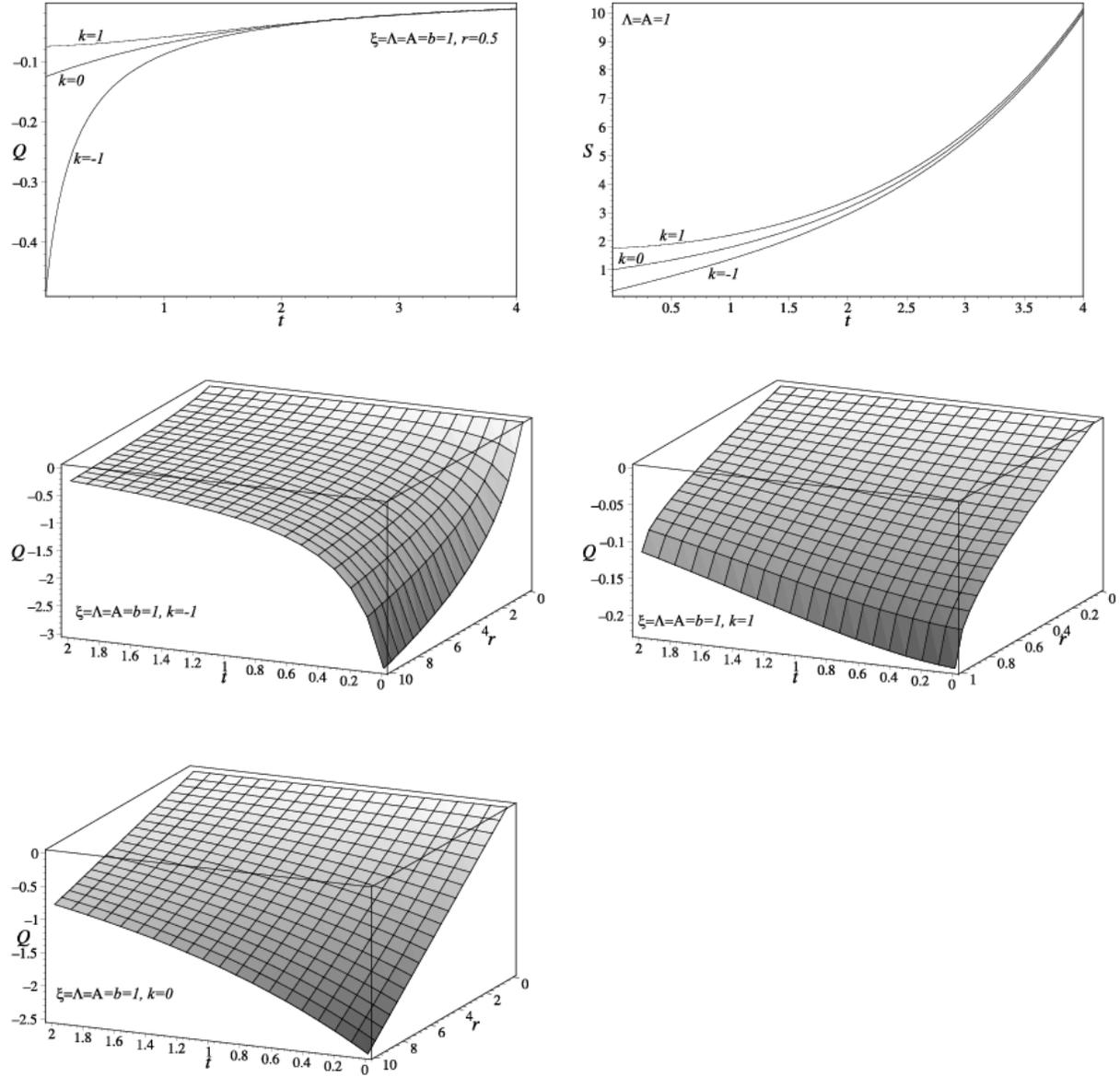}
    \caption[Vacuum solution]{Scale factor $S$ and the function $Q$ (which governs
the Weyl 1-form) in case of the vacuum ($\Lambda \neq 0$, $\chi = 0$)
solution.}
     \label{figVacuumSolution}
   \end{center}
\end{figure}

\section{Intermediate vacuum solution\label{INTERMEDIATE_SOLUTION_KAPITEL}}

At this point we will briefly mention a solution in case of $\chi =\Lambda
=0\neq k$. Under this assumption eq. (\ref{2_final}) is fulfilled
identically. Equation (\ref{1_final}) and (\ref{3_final}) turn into 
\begin{eqnarray}
\frac{\ddot{S}}{S}+\left( \frac{\dot{S}}{S}\right) ^{2}+\frac{k}{S^{2}} &=&0,
\label{1_intermediate_vacuum} \\
-\frac{\kappa \mu _{0}}{3S^{4}} &=&0.  \label{3_intermediate_vacuum}
\end{eqnarray}
It is straightforward to check that\footnote{$[F]=$length$^{2}$.} 
\begin{equation}
S(t)=\sqrt{F-k\,t^{2}},\texttt{\rm \quad \quad with\quad }F=\texttt{\rm const,}
\label{intermediate_scale_factor}
\end{equation}
is a solution of (\ref{1_intermediate_vacuum}). From (\ref
{3_intermediate_vacuum}) we recover the same relation between the energy and
stresses as in (\ref{stresses_and_energy_for_vacuum_case_final}). Thus, this
solution belongs to the vacuum regime, too. The final form of the Weyl
1-form reads 
\begin{eqnarray}
Q &=-\frac{\xi (t)}{4b\sqrt{F+t^{2}}}\,\,\texttt{\rm arcsinh}(r)\,\vartheta ^{%
\hat{0}}\quad \quad &\texttt{\rm for }k=-1,
\label{weyl_final_intermediate_k_minus_1} \\
Q &=-\frac{\xi (t)}{4b\sqrt{F-t^{2}}}\,\,\texttt{\rm arcsin}(r)\,\vartheta ^{\hat{%
0}}\quad \quad &\texttt{\rm for }k=1.  \label{weyl_final_intermediate_k_1}
\end{eqnarray}
The surviving curvature components are given by 
\begin{eqnarray}
^{(4)}Z_{\hat{0}\hat{0}} &=&\frac{\xi (t)}{8b\left( F-kt^{2}\right) }%
\vartheta ^{\hat{0}}\wedge \vartheta ^{\hat{1}}=-\,^{(4)}Z_{\hat{1}\hat{1}%
}=-\,^{(4)}Z_{\hat{2}\hat{2}}=-\,^{(4)}Z_{\hat{3}\hat{3}}\texttt{\rm ,} \\
^{(4)}W^{\alpha \beta } &=&\frac{kF}{\left( F-kt^{2}\right) ^{2}}\vartheta
^{\alpha }\wedge \vartheta ^{\beta }\texttt{\rm ,}
\end{eqnarray}
The parameter values and epochs at which the curvature invariant diverges
are collected in table \ref{tabelle_3}.
\begin{table}
\caption{Parameter values and times with essential singularities (intermediate solution).}
\label{tabelle_3}
\begin{indented}
\item[]
\begin{tabular}{@{}ll}
\br
$k=-1$ & $\{b=0\}$ or $\{t_{{\rm div}}=\pm \sqrt{-F}\}$ \\
$k=1$ & $\{b=0\}\,$or $\{t_{{\rm div}}=\pm \sqrt{F}\}$ \\ 
\br
\end{tabular}
\end{indented}
\end{table}
Again, these times correspond to a point singularity at the origin of the
universe, i.e. $S(t_{{\rm div}})=0$. 
\begin{figure}
  \begin{center}
    \epsfig{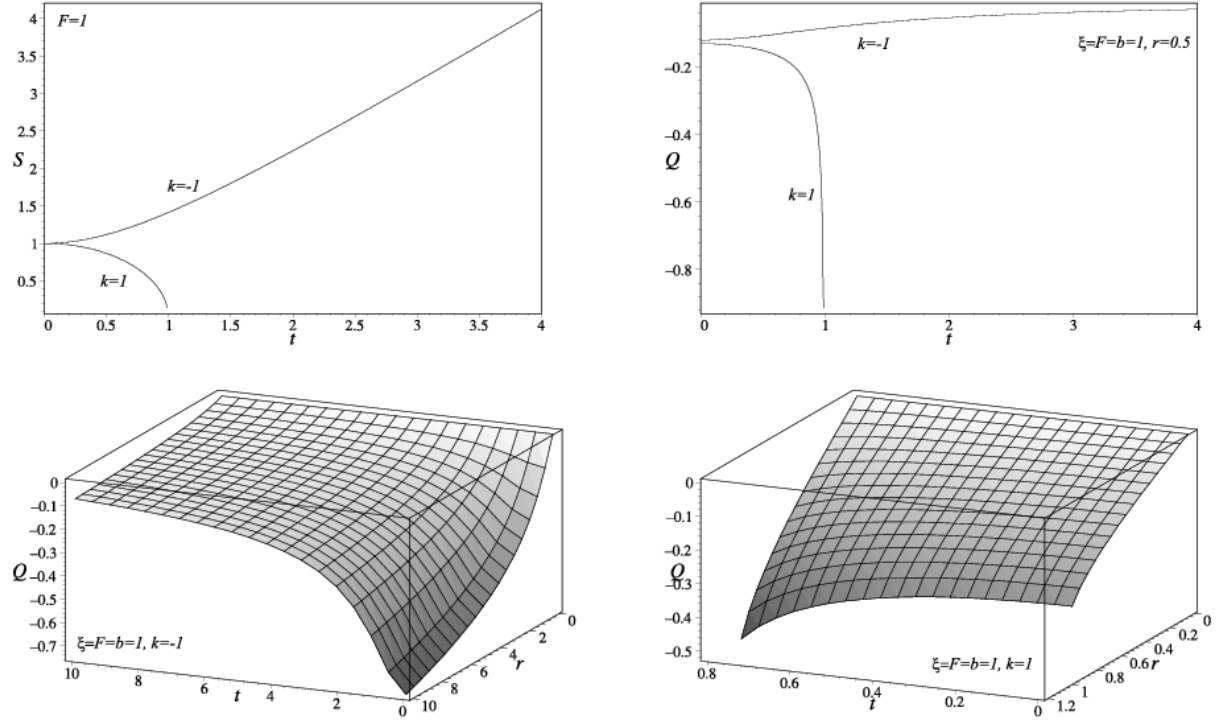}
    \caption[Intermediate solution]{Scale factor and the function $Q$ (which governs
the Weyl 1-form) in case of the intermediate solution, i.e. $\Lambda = \chi =
0$.}
     \label{figIntermediateSolution}
   \end{center}
\end{figure}

\section{Radiative solution\label{RADIATIVE_SOLUTION_KAPITEL}}

Let us now investigate the branch $\chi \neq 0$, i.e.\thinspace the induced
cosmological constant $\Lambda $ is forced to vanish because of (\ref
{2_final}). The remaining two field equations (\ref{1_final}) and (\ref
{3_final}) turn into 
\begin{eqnarray}
\frac{\ddot{S}}{S}+\left( \frac{\dot{S}}{S}\right) ^{2}+\frac{k}{S^{2}} &=&0,
\label{final_1_lambda_null} \\
\chi \frac{\ddot{S}}{S}+\frac{\kappa \mu _{0}}{3S^{4}} &=&0.
\label{final_3_lambda_null}
\end{eqnarray}
This set is solved by 
\begin{eqnarray}
S(t) &=\sqrt{\frac{\kappa \mu _{0}}{3k\chi }-kt^{2}}\quad \quad &\texttt{\rm for }%
k=-1,1\,,  \label{radiative_scale_factor_1} \\
S(t) &=\sqrt{C+2\sqrt{\frac{\kappa \mu _{0}}{3\chi }}t} \quad \quad &\texttt{\rm for }k=0.  \label{radiative_scale_factor_2}
\end{eqnarray}
Here $C$ denotes an arbitrary constant\footnote{$[C]=$length$^{2}$.}. Again,
we derive the effective energy and pressures (\ref
{effective_energy_and_pressure_definition}): 
\begin{eqnarray}
^{{\rm rad}}\tilde{\mu}:= &&\,^{{\rm rad}}\mu -\,^{{\rm vac}}\mu =\frac{%
\mu _{0}}{S^{4}},\quad  \label{pressure_and_energy_for_radiative_solution_1}
\\
^{{\rm rad}}\tilde{p}_{r}:= &&\,^{{\rm rad}}p_{r}-\,^{{\rm vac}}p_{r}=%
\frac{\mu _{0}}{3S^{4}},\quad ^{{\rm rad}}\tilde{p}_{t}:=\,^{{\rm rad}%
}p_{t}-\,^{{\rm vac}}p_{t}=\frac{\mu _{0}}{3S^{4}}.
\label{pressure_and_energy_for_radiative_solution_2}
\end{eqnarray}
Thus, with $\,^{{\rm rad}}\tilde{p}:=\,^{{\rm rad}}\tilde{p}_{r}=\,^{{\rm %
rad}}\tilde{p}_{t}$, the radiative relation between the effective energy and
pressure holds: 
\begin{equation}
^{{\rm rad}}\tilde{p}=\frac{^{{\rm rad}}\tilde{\mu}}{3}\,.
\end{equation}
Finally, the Weyl 1-form reads 
\begin{eqnarray}
Q &=-\frac{\xi (t)}{4b\sqrt{t^{2}-\frac{\kappa \mu _{0}}{3\chi }}}\,\texttt{\rm %
arcsinh}(r)\,\vartheta ^{\hat{0}}\quad \quad &\texttt{\rm for }k=-1,
\label{weyl_radiative_k_minus} \\
Q &=-\frac{\xi (t)\,r}{4b\sqrt{C+2\sqrt{\frac{\kappa \mu _{0}}{3\chi }}t}}%
\,\vartheta ^{\hat{0}}\quad \quad &\texttt{\rm for }k=0,
\label{weyl_radiative_k_null} \\
Q &=-\frac{\xi (t)}{4b\sqrt{\frac{\kappa \mu _{0}}{3\chi }-t^{2}}}\,\texttt{\rm %
arcsin}(r)\,\vartheta ^{\hat{0}}\quad \quad &\texttt{\rm for }k=1.
\label{weyl_radiative_k_plus}
\end{eqnarray}
\newline
Again, we observe that the only surviving pieces of the curvature are $%
^{(4)}W_{\alpha \beta }\,$\ and $^{(4)}Z_{\alpha \beta }$, their
nonvanishing components are given by: 
\begin{eqnarray}
^{(4)}Z_{\hat{0}\hat{0}} &=\frac{3\chi k\xi (t)}{8b\left( 3\chi
t^{2}-\kappa \mu _{0}\right) }\,\vartheta ^{\hat{0}}\wedge \vartheta ^{\hat{1%
}} & \nonumber \\
&=-\,^{(4)}Z_{\hat{1}\hat{1}}=-\,^{(4)}Z_{\hat{2}\hat{2}}=-\,^{(4)}Z_{\hat{3%
}\hat{3}}\,&\texttt{\rm for }k=1,-1, \\
^{(4)}Z_{\hat{0}\hat{0}} &=\frac{\xi (t)}{8b\left( 2\sqrt{\frac{\kappa \mu
_{0}}{3\chi }}t+C\right) }\,\vartheta ^{\hat{0}}\wedge \vartheta ^{\hat{1}} 
&\nonumber \\
&=-\,^{(4)}Z_{\hat{1}\hat{1}}=-\,^{(4)}Z_{\hat{2}\hat{2}}=-\,^{(4)}Z_{\hat{3%
}\hat{3}}\,&\texttt{\rm for }k=0, \\
^{(4)}W^{\alpha \beta } &=\frac{3\chi \kappa \mu _{0}}{\left( 3\chi
t^{2}-\mu _{0}\kappa \right) ^{2}}\vartheta ^{\alpha }\wedge \vartheta
^{\beta }\,\quad &\texttt{\rm for }k=-1,1\,,
\label{rotational_curvature_radiative_k_1_minus_1} \\
^{(4)}W^{\alpha \beta } &=\frac{\kappa \mu _{0}}{12\sqrt{\frac{\kappa \mu
_{0}}{3\chi }}\chi Ct+3\chi C^{2}+4\kappa \mu _{0}t^{2}}\,\vartheta ^{\alpha
}\wedge \vartheta ^{\beta }\quad &\texttt{\rm for }k=0.
\label{rotational_curvature_radiative_k_zero}
\end{eqnarray}
In table \ref{tabelle_4} we list the parameter values and epochs at which the
curvature invariant diverges.
\begin{table}
\caption{Parameter and times with essential singularities (radiative solution).}
\label{tabelle_4}
\begin{indented}
\item[]\begin{tabular}{@{}ll}
\br
$k=-1$ & $\{b=0\}\,\,$or$\,\,\,\{t_{{\rm div}}=\pm \sqrt{\frac{\mu
_{0}\kappa }{3\chi }}\}$ or $\{\kappa =0,\,\chi =0\}\,$\ or $\{t=0,\,\kappa
=0\}$ \\ 
$k=0$ & $\{b=0\}$ or$\,\,\,\{t_{{\rm div}}=-\frac{C}{2}\sqrt{\frac{3\chi }{%
\kappa \mu _{0}}}\}$ \\
$k=1$ & $\{b=0\}\,\,$or$\,\,\,\{t_{{\rm div}}=\pm \sqrt{\frac{\mu
_{0}\kappa }{3\chi }}\}$ or $\{\kappa =0,\,\chi =0\}\,$\ or $\{t=0,\,\kappa
=0\}$ \\ 
\br
\end{tabular}
\end{indented}
\end{table}

The situation is the same as in case of the vacuum and intermediate
solution, i.e. the times $t_{{\rm div}}$ correspond to
$S(t_{{\rm div}})=0$, yielding a point singularity at the origin of the
universe. 
\begin{figure}
  \begin{center}
    \epsfig{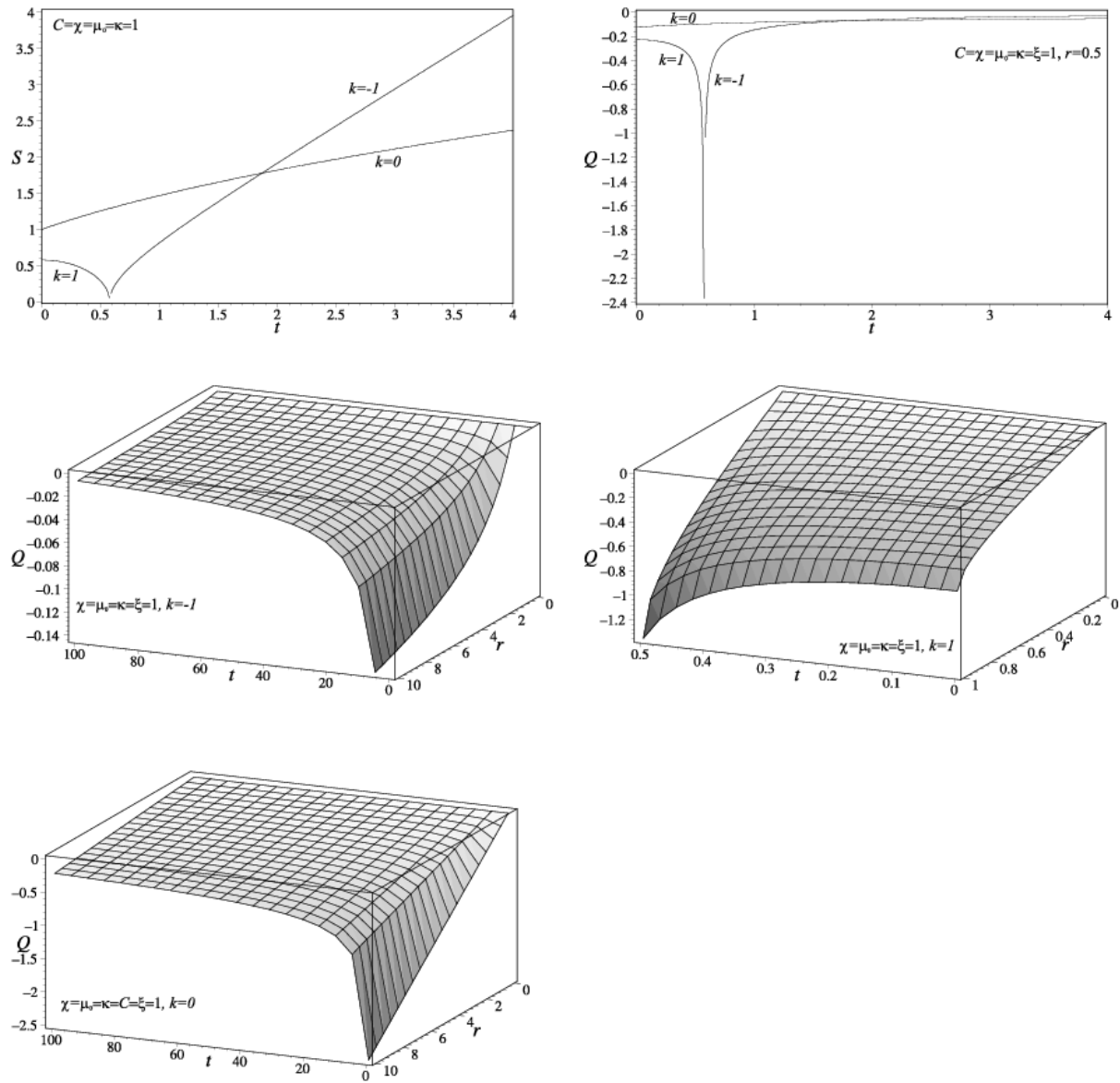}
    \caption[Radiative solution]{Scale factor $S$ and the function $Q$ (which governs
the Weyl 1-form) in case of the radiative ($\Lambda = 0$, $\chi \neq 0$)
solution.}
     \label{figRadiativeSolution}
   \end{center}
\end{figure}

\section{Conclusion}

We were able to recover the results found in \cite{Tresguerres}. We fixed some dimensional problems and calculated the
explicit form of the curvature pieces for all branches of the solution. From
the figures \ref{figVacuumSolution}--\ref{figRadiativeSolution} we gain
insight into the behavior of the Weyl 1-from $Q$ which controls the
non-Riemannian features of the solution. Since we expect that the
non-Riemannian quantities are only present at very early stages of the
universe, the explicit form of $Q$ (cf. eqs. (\ref{weyl_vacuum_k_minus})-(\ref{weyl_vacuum_k_plus}), (\ref{weyl_final_intermediate_k_minus_1})-(\ref
{weyl_final_intermediate_k_1}), (\ref{weyl_radiative_k_minus})-(\ref
{weyl_radiative_k_plus})) restricts the allowed choices for the function $
\xi (t)$ entering our ansatz in (\ref{3_ansatz_polarisation_form}). Thus, in
case of a suitable choice for $\xi (t)$, the solution asymptotically evolves
from a Weyl-Cartan into a Riemannian spacetime. We note that some of the
previously found branches of the solution can be matched. For this purpose,
we investigate the $k=-1$ branch. The corresponding scale factors are
listed in table \ref{tabelle_5}.
\begin{table}
\caption{Comparison of the scale factors in case of $k=-1$ for all branches.}
\label{tabelle_5}
\begin{indented}
\item[]\begin{tabular}{@{}llll}
\br
 &Vacuum & Intermediate & Radiative \\ 
\mr
Parameters & $\Lambda \neq 0,\chi =0$ & $\Lambda =0,\chi =0$ & $\Lambda =0,\chi \neq 0$
\\ 
Scale factors & $S(t)=A\,e^{\sqrt{\frac{\Lambda }{3}}t}-\frac{3}{4A\Lambda }\,e^{-\sqrt{%
\frac{\Lambda }{3}}t}$ & $S(t)=\sqrt{F+\,t^{2}}$ & $S(t)=\sqrt{t^{2}-\frac{%
\kappa \mu _{0}}{3\chi }}$ \\ 
\br
\end{tabular}
\end{indented}
\end{table}
From this table it is clear that the intermediate and radiative scale
factors can be matched by choosing $F=-\frac{\kappa \mu _{0}}{3\chi }$.

\begin{figure}
  \begin{center}
    \epsfig{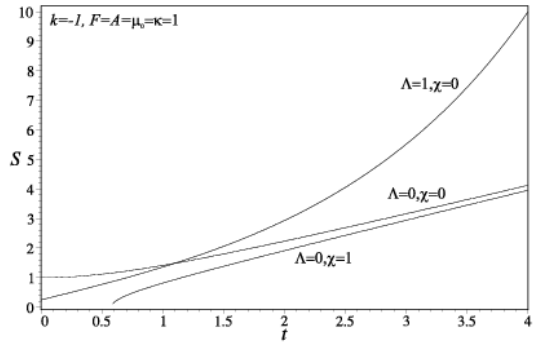}
   \caption[Scale factors for all branches]{Scale factors for all three
branches of the solution, i.e. the vacuum ($\Lambda \neq 0$, $\chi = 0$),
intermediate ($\Lambda = \chi = 0$), and radiative ($\Lambda = 0$, $\chi \neq
0$) part.}
     \label{figAllScaleFactors}
   \end{center}
\end{figure}

In figure \ref{figAllScaleFactors} we display all three branches in one
plot. With the exception of $t$, we set all free parameters to the value $1$. 
Finally, we note that the singularities of the solution are of the same type
as in the Riemannian case, i.e. point singularities at the origin of the universe.
This result is rather surprising, because most of the known extensions of
the cosmological standard model exhibit singularities of extended
geometrical shape (cf. \cite{Boerner}, \cite{MacCallum}) or, in some cases,
lead to the avoidance of a singularity (cf. \cite{TuckerDark}). In future work it
might be interesting to incorporate anisotropic and inhomogeneous metrical
structures into our model.


\appendix

\section{MAG in general\label{MAG_KAPITEL}}

In MAG we have the metric $g_{\alpha \beta }$, the coframe $\vartheta
^{\alpha }$, and the connection 1-form $\Gamma _{\alpha}{}^{\beta}$ (with
values in the Lie algebra of the four-dimensional linear group $GL(4,R)$) as
new independent field variables.\ Here $\alpha ,\beta ,\ldots =0,1,2,3$
denote (anholonomic) frame indices. Spacetime is described by a
metric-affine geometry with the gravitational field strengths nonmetricity $%
Q_{\alpha \beta }:=-Dg_{\alpha \beta }$, torsion $T^{\alpha }:=D\vartheta
^{\alpha }$, and curvature $R_{\alpha }{}^{\beta }:=d\Gamma _{\alpha
}{}^{\beta }-\Gamma _{\alpha }\,^{\gamma }\wedge \Gamma _{\gamma }{}^{\beta
} $. A Lagrangian formalism for a matter field $\Psi $ minimally coupled to
the gravitational potentials $g_{\alpha \beta }$, $\vartheta ^{\alpha }$, $%
\Gamma _{\alpha }{}^{\beta }$ has been set up in \cite{PhysRep}. 
An alternative interpretation of the metric as a Goldstone field, no
more playing the role of a fundamental gravitational potential, has been
proposed by Mielke and one of the authors in \cite{Mielke} in the context of
nonlinear realizations of spacetime groups. The dynamics of an ordinary MAG theory is specified by a total Lagrangian 
\begin{equation}
L=V_{{\rm MAG}}(g_{\alpha \beta },\vartheta ^{\alpha },Q_{\alpha \beta
},T^{\alpha },R_{\alpha }{}^{\beta })+L_{{\rm mat}}(g_{\alpha \beta
},\vartheta ^{\alpha },\Psi ,D\Psi ).
\end{equation}
The variation of the action with respect to the independent gauge potentials
leads to the field equations: 
\begin{eqnarray}
\frac{\delta L_{{\rm mat}}}{\delta \Psi } &=&0,  \label{matter} \\
DM^{\alpha \beta }-m^{\alpha \beta } &=&\sigma ^{\alpha \beta },
\label{zeroth} \\
DH_{\alpha }-E_{\alpha } &=&\Sigma _{\alpha ,}  \label{first} \\
DH^{\alpha }{}_{\beta }-E^{\alpha }{}_{\beta } &=&\Delta ^{\alpha }{}_{\beta
}.  \label{second}
\end{eqnarray}
Equation (\ref{first}) is the generalized Einstein equation with the
energy-momentum 3-form $\Sigma _{\alpha }$ as its source, whereas (\ref
{zeroth}) and (\ref{second}) are additional field equations which take into
account other aspects of matter, such as spin, shear and dilation currents,
represented by the hypermomentum $\Delta ^{\alpha }{}_{\beta }$. We made use
of the definitions of the gauge field excitations, 
\begin{equation}
H_{\alpha }:=-\frac{\partial V_{{\rm MAG}}}{\partial T^{\alpha }},\quad
H^{\alpha }{}_{\beta }:=-\frac{\partial V_{{\rm MAG}}}{\partial R_{\alpha
}{}^{\beta }},\quad M^{\alpha \beta }:=-2\frac{\partial V_{{\rm MAG}}}{%
\partial Q_{\alpha \beta }},  \label{exications}
\end{equation}
of the canonical energy-momentum, the metric stress-energy and the
hypermomentum current of the gauge fields, 
\begin{equation}
E_{\alpha }:=\frac{\partial V_{{\rm MAG}}}{\partial \vartheta ^{\alpha }}%
,\quad m^{\alpha \beta }:=2\frac{\partial V_{{\rm MAG}}}{\partial g_{\alpha
\beta }},\quad E^{\alpha }{}_{\beta }:=-\vartheta ^{\alpha }\wedge H_{\beta
}-g_{\beta \gamma }M^{\alpha \gamma },  \label{gauge_currents}
\end{equation}
and of the canonical energy-momentum, the metric stress-energy and the
hypermomentum currents of the matter fields, 
\begin{equation}
\Sigma _{\alpha }:=\frac{\delta L_{{\rm mat}}}{\delta \vartheta ^{\alpha }}%
,\quad \sigma ^{\alpha \beta }:=2\frac{\delta L_{{\rm mat}}}{\delta
g_{\alpha \beta }},\quad \Delta ^{\alpha }{}_{\beta }:=\frac{\delta L_{{\rm %
mat}}}{\delta \Gamma _{\alpha }{}^{\beta }}.  \label{matter_currents}
\end{equation}
Provided the matter equations (\ref{matter}) are fulfilled, the following
Noether identities hold: 
\begin{eqnarray}
D\Sigma _{\alpha } &=&\left( e_{\alpha }\rfloor T^{\beta }\right) \wedge
\Sigma _{\beta }-\frac{1}{2}\left( e_{\alpha }\rfloor Q_{\beta \gamma
}\right) \sigma ^{\beta \gamma }+\left( e_{\alpha }\rfloor R_{\beta
}{}^{\gamma }\right) \wedge \Delta ^{\beta }{}_{\gamma },
\label{noether_ident_1} \\
D\Delta ^{\alpha }{}_{\beta } &=&g_{\beta \gamma }\sigma ^{\alpha \gamma
}-\vartheta ^{\alpha }\wedge \Sigma _{\beta }.  \label{noether_ident_2}
\end{eqnarray}
They show that the field equation (\ref{zeroth}) is redundant, thus we only
need to take into account (\ref{first}) and (\ref{second}).

As suggested in \cite{PhysRep}, the most general parity conserving quadratic
Lagrangian expressed in terms of the irreducible pieces of the nonmetricity $%
Q_{\alpha \beta }$, torsion $T^{\alpha }$, and curvature $R_{\alpha \beta }$
reads 
\begin{eqnarray}
V_{{\rm MAG}}=&\frac{1}{2\kappa }\biggl[ &-a_{0}R^{\alpha \beta }\wedge \eta
_{\alpha \beta }-2\lambda \eta +T^{\alpha }\wedge \,^{\star }\left(
\sum_{I=1}^{3}a_{I}\,^{(I)}T_{\alpha }\right)   \nonumber \\
&&+Q_{\alpha \beta }\wedge \,^{\star }\left(
\sum_{I=1}^{4}b_{I}\,^{(I)}Q^{\alpha \beta }\right)\nonumber \\  &&+b_{5}\left(
^{(3)}Q_{\alpha \gamma }\wedge \vartheta ^{\alpha }\right) \wedge \,^{\star
}\left( \,^{(4)}Q^{\beta \gamma }\wedge \vartheta _{\beta }\,\right)   \nonumber
\\
&&+2\left( \sum_{I=2}^{4}c_{I}\,^{(I)}Q_{\alpha \beta }\right) \wedge
\vartheta ^{\alpha }\wedge \,^{\star }T^{\beta }\biggr]  \nonumber \\
-\frac{1}{2\rho }\, R^{\alpha \beta }\wedge \,^{\star
}&&\hspace{-0.5cm}\biggl[\sum_{I=1}^{6}w_{I}\,^{(I)}W_{\alpha \beta
}+\sum_{I=1}^{5}z_{I}\,^{(I)}Z_{\alpha \beta }+w_{7}\vartheta _{\alpha
}\wedge \left( e_{\gamma }\rfloor \,^{(5)}W^{\gamma }{}_{\beta }\right)  
\nonumber \\
&&\hspace{-0.5cm}+z_{6}\vartheta _{\gamma }\wedge \left( e_{\alpha }\rfloor
\,^{(2)}Z^{\gamma }{}_{\beta }\right) +\sum_{I=7}^{9}z_{I}\vartheta _{\alpha
}\wedge \left( e_{\gamma }\rfloor \,^{(I-4)}Z^{\gamma }{}_{\beta }\right) \biggr].
\label{general_v_mag}
\end{eqnarray}
The constants entering (\ref{general_v_mag}) are the cosmological constant $%
\lambda $, the weak and strong coupling constant $\kappa $ and $\rho $%
\footnote{$[\lambda ]=$length$^{-2}$, $[\kappa ]=$length$^{2}$, $[\rho
]=[\hbar ]=[c]=1.$}, and the 28 dimensionless parameters 
\begin{equation}
a_{0},\dots ,a_{3},b_{1},\dots ,b_{5},c_{2},\dots ,c_{4},w_{1},\dots
,w_{7},z_{1},\dots ,z_{9}.  \label{general_coupling}
\end{equation}
This Lagrangian and the presently known exact solutions in MAG have been
reviewed in \cite{Exact2}. We note that this Lagrangian incorporates the one
used in section \ref{TRESGUERRES_KAPITEL} eq. (\ref{Tresguerres_lagrangian}%
), as can be seen easily by making the following choice for the constants in
(\ref{general_v_mag}): 
\begin{equation}
\lambda ,a_{1},\dots ,a_{3},b_{1},\dots ,b_{5},c_{2},\dots
,c_{4},w_{7},z_{1},\dots ,z_{3},z_{5},\dots ,z_{9}=0.\quad 
\end{equation}
In order to obtain exactly the form of (\ref{Tresguerres_lagrangian}), one
has to perform the additional substitutions: 
\begin{equation}
a_{0}\rightarrow -\chi ,\quad w_{1},\dots ,w_{6}\rightarrow -2\rho
a_{1},\dots ,-2\rho a_{6},\quad z_{4}\rightarrow -2\rho b.
\end{equation}

\section{Weyl-Cartan spacetime\label{WEYL_CARTAN_KAPITEL}}

The Weyl-Cartan $\,$spacetime ($Y_{n}$) is a special case of the general
metric-affine geometry in which the tracefree part $Q_{\alpha \beta
  }\!\!\!\!\!\!\!\!\!\!\!\!\!\nearrow \,\,\,\,\,\,\,$ of the nonmetricity
$Q_{\alpha \beta }\,$ vanishes. Thus, the whole
nonmetricity is proportional to its trace part, i.e. the Weyl 1-form $Q:=%
\frac{1}{4}Q^{\alpha }{}_{\alpha }$, 
\begin{equation}
Q_{\alpha \beta }=g_{\alpha \beta }\,Q=\frac{1}{4}g_{\alpha \beta
}\,Q^{\gamma }{}_{\gamma }.  \label{torsion_in_weyl_cartan_spacetime}
\end{equation}
Therefore the general MAG connection reduces to 
\begin{eqnarray}
\Gamma _{\alpha \beta } &=&\frac{1}{2}dg_{\alpha \beta }+\left( e_{[\alpha
}\rfloor dg_{\beta ]\gamma }\right) \vartheta ^{\gamma }+e_{[\alpha }\rfloor
C_{\beta ]}-\frac{1}{2}\left( e_{\alpha }\rfloor e_{\beta }\rfloor C_{\gamma
}\right) \vartheta ^{\gamma }  \nonumber \\
&&-e_{[\alpha }\rfloor T_{\beta ]}+\frac{1}{2}\left( e_{\alpha }\rfloor
e_{\beta }\rfloor T_{\gamma }\right) \vartheta ^{\gamma }+\frac{1}{2}%
g_{\alpha \beta }\,Q+\left( e_{[\alpha }\rfloor Q\right) \vartheta _{\beta ]}
\\
&=&\Gamma _{\alpha \beta }^{\{\,\}}-e_{[\alpha }\rfloor T_{\beta ]}+\frac{1}{%
2}\left( e_{\alpha }\rfloor e_{\beta }\rfloor T_{\gamma }\right) \vartheta
^{\gamma }+\frac{1}{2}g_{\alpha \beta }\,Q+\left( e_{[\alpha }\rfloor
Q\right) \vartheta _{\beta ]}.  \label{Weyl_Cartan_Konnektion}
\end{eqnarray}
Thus, it not longer includes a symmetric tracefree part. Now let us
recall the definition of the material hypermomentum $\Delta _{\alpha \beta }$
given in (\ref{matter_currents}). Due to the absence of a symmetric
tracefree piece in (\ref{Weyl_Cartan_Konnektion}), $\Delta _{\alpha \beta }$
decomposes as follows 
\begin{eqnarray}
\Delta _{\alpha \beta } &=&\texttt{\rm antisymmetric piece + trace piece}  \nonumber
\\
&=&\tau _{\alpha \beta }+\frac{1}{4}g_{\alpha \beta }\,\Delta =\tau _{\alpha
\beta }+\frac{1}{4}g_{\alpha \beta }\,\Delta ^{\gamma }{}_{\gamma }  \nonumber
\\
&=&\texttt{\rm spin current + dilation current.}  \label{Weyl_Cartan_hypermomentum}
\end{eqnarray}
According to (\ref{Weyl_Cartan_hypermomentum}) the second MAG field equation
(\ref{second}) decomposes into 
\begin{eqnarray}
dH^{\alpha }{}_{\alpha }-E^{\alpha }{}_{\alpha } &=&\Delta , \\
g_{\gamma \lbrack \alpha }DH^{\gamma }\,_{\beta ]}-E_{[\alpha \beta ]}
&=&\tau _{\alpha \beta },
\end{eqnarray}
while the first field equation is still given by (\ref{first}).
Additionally, we can decompose the second Noether identity (\ref
{noether_ident_2}) into 
\begin{eqnarray}
\frac{1}{4}g_{\alpha \beta }\,d\Delta +\vartheta _{(\alpha }\wedge \Sigma
_{\beta )} &=&\sigma _{\alpha \beta },  \label{weyl_cartan_second_noether_1}
\\
D\tau _{\alpha \beta }+Q\wedge \tau _{\alpha \beta }+\vartheta _{\lbrack
\alpha }\wedge \Sigma _{\beta ]} &=&0.  \label{weyl_cartan_second_noether_2}
\end{eqnarray}
Thus, the first Noether identity (\ref{noether_ident_1}) with inserted Weyl
1-form and hypermomentum reads 
\begin{equation}
D\Sigma _{\alpha }=\left( e_{\alpha }\rfloor T^{\beta }\right) \wedge \Sigma
_{\beta }-\frac{1}{2}\left( e_{\alpha }\rfloor Q\right) \sigma ^{\beta
}{}_{\beta }+\left( e_{\alpha }\rfloor R_{[\beta \gamma ]}{}\right) \wedge
\tau ^{\beta \gamma }+\frac{1}{4}\left( e_{\alpha }\rfloor R\right) \wedge
\Delta .  \label{connection_weyl_cartan_final}
\end{equation}
Finally, we note that in a $Y_{n}$ spacetime the symmetric part of the
curvature $R_{(\alpha \beta )}=Z_{\alpha \beta }\,$, i.e. the strain
curvature, reduces to the trace part 
\begin{equation}
Z_{\alpha \beta }=\frac{1}{4}g_{\alpha \beta }R=\frac{1}{4}g_{\alpha \beta
}R^{\gamma }{}_{\gamma }=\frac{1}{2}g_{\alpha \beta }\,dQ.
\end{equation}

\section{Units}

In this work we made use of \textit{natural units}, i.e. $\hbar =c=1$
(cf. table \ref{tabelle_6}).
\begin{table}
\caption{Natural units.}
\label{tabelle_6}
\begin{indented}
\item[]\begin{tabular}{@{}llll}
\br
[energy] & [mass] & [time] & [length] \\ 
\mr
length$^{-1}$ & length$^{-1}$ & length & length \\ 
\br
\end{tabular}
\end{indented}
\end{table}
Additionally, we have to be careful with the coupling constants and the
coordinates within the coframe. In order to keep things as clear as
possible, we provide a list of the quantities emerging in section \ref
{TRESGUERRES_KAPITEL} in table \ref{tabelle_7}.
\begin{table}
\caption{Units of quantities in section \ref{TRESGUERRES_KAPITEL}.} 
\label{tabelle_7}
\begin{indented}
\item[]\begin{tabular}{@{}ll}
\br
Quantities & Units \\ \mr
Coordinates & $[t]=$ length, $[\phi ]=[r]=[\theta ]=1$ \\ 
Constants & $[k]=[\chi ]=[b]=1,$ $[\Lambda ]^{-\frac{1}{2}}=[\kappa ]^{\frac{%
1}{2}}=$ length \\ 
Functions & $[S(t)]=$ length, $[\xi (t)]=1$ \\ 
Additional constants & $[A]=[F]^{\frac{1}{2}}=[C]^{\frac{1}{2}}=$ length, $%
[\mu _{0}]=1$ \\ 
\br
\end{tabular}
\end{indented}
\end{table}
Note that $[d]=1$ and $[\,^{\star }]=$ length$^{n-2p}$, where $n=$ dimension
of the spacetime, $p=$ degree of the differential form on which $^{\star }$
acts.

\ack
The authors are grateful to Prof. F.W. Hehl, and the members of the gravity
group at the University of Cologne for their support. 

\end{document}